\documentclass[prl,aps,twocolumn,showpacs]{revtex4}
\begin{document}
\title{Z$_3$ Quantum Criticality in a spin-1/2 chain model}
\author{P. Lecheminant}
\affiliation{Laboratoire de Physique Th\'eorique et
Mod\'elisation, CNRS UMR 8089,
Universit\'e de Cergy-Pontoise, 5 Mail Gay-Lussac, Neuville sur Oise,
95031 Cergy-Pontoise Cedex, France}
\author{E. Orignac}
\affiliation{Laboratoire de Physique Th\'eorique, CNRS UMR 8549,
Ecole Normale Sup{\'e}rieure, 24 rue Lhomond, 75231 Paris Cedex 05, France}
\begin{abstract}
The stability of the magnetization $m=1/3$ plateau phase of
the XXZ spin-1/2 Heisenberg chain
with competing interactions is investigated 
upon switching on a staggered transverse
magnetic field. 
Within a bosonization approach, 
it is shown that the low-energy properties of the model
are described by an effective two-dimensional
XY model in a three-fold symmetry-breaking field.
A phase transition in the three-state Potts universality
class is expected separating the $m=1/3$ plateau phase
to a phase where the spins are polarized along the 
staggered magnetic field.
The Z$_3$ critical properties of the transition are
determined within the bosonization approach.
\end{abstract}            
\maketitle

Quantum phase transitions i.e. zero-temperature 
transitions  driven entirely by quantum fluctuations  have attracted much interest 
over the past decade \cite{sachdevbook}. 
Quantum criticality can emerge from the interplay 
of  competing orders \cite{sachdev00,zhang97}, and
the existence of a quantum critical point offers firm ground
to develop a controlled theory of the intermediate
regimes characterized by multiple competing orders \cite{sachdev00}.
In one-dimensional quantum systems, a full description 
of quantum critical points can be obtained  
based on powerful analytical and numerical techniques 
such as conformal field theory (CFT) and integrability for the former 
and density matrix renormalization group (DMRG) \cite{white} and exact
diagonalizations for the latter. 
In particular, the interplay between different kind of
orders can be described, in an analytic continuum approach, 
by a CFT perturbed by several relevant terms (in the
renormalization group sense).
When acting separately, each perturbation yields a strong-coupling
massive ordered phase but the competition between them can give rise
to a quantum critical point at intermediate coupling.
Recently, several examples of this behavior have been analyzed
non-perturbatively, such as the emergence of an Ising (or Z$_2$) critical point
in the two-frequency sine-Gordon model \cite{delfino98,fgn99,bajnok01}
or in the $\beta^2 = 4 \pi$ self-dual sine-Gordon (SDSG) model
\cite{ogilvie81,lgn02}. This Z$_2$ quantum criticality has been found 
to be relevant for various spin chains
\cite{fgn99,schulz86,totsuka98,ye01,wang02,alergo02}
or  one-dimensional  interacting fermion models
\cite{fgn99,edmond01,furusaki02}. 
Richer patterns of quantum criticalities are expected to be 
realized in more complex situations.
In this respect, U(1) Gaussian criticality and SU(2)$_k$ ($k=1,2$)
Wess-Zumino-Novikov-Witten universality classes have
been found in the two-leg spin ladder with external perturbations
such as a staggered magnetic field \cite{wang02}, an explicit dimerization
\cite{affleck87,delgado96,cabra99,wang00}, 
and biquadratic interactions \cite{nersesyan97,nomura02}. However, all
of these critical points can still be described in terms of free bosons or
free Majorana fermions. An interesting question is whether a lattice model with
only short-range interactions can yield more complex criticality. 

In this letter, we shall discuss a XXZ spin-1/2 Heisenberg chain with
competing interactions in magnetic fields which exhibits 
an emerging quantum criticality
in the three-state Potts (or Z$_3$) universality class.
We note that a similar critical behavior has  been
obtained previously in a numerical investigation of a quantum spin chain model
 in transverse and longitudinal 
magnetic fields \cite{penson88}. However, this latter model 
contained a three-spin interaction  whereas the
model, considered here, involves only pairwise interactions. 
The Hamiltonian of our model reads as follows:
\begin{eqnarray}
{\cal H} &=& J \sum_n {\bf S}_n \cdot {\bf S}_{n+1}
+ \Delta_1 \sum_n S_n^z
 S_{n+1}^z   +  \Delta_2 \sum_n S_n^z
  S_{n+2}^z \nonumber \\
&-& H \sum_n S_n^z + h_s \sum_n (-1)^n S_n^x,
 \label{latticemodel}
\end{eqnarray}
where $S_n^{a}$, $a=x,y,z$ is a spin-1/2 operator
at site $n$ and $\Delta_{1,2},J > 0$.
In absence of the staggered magnetic field $h_s = 0$,
our model corresponds to a
frustrated XXZ Heisenberg chain with easy axis anisotropy in a magnetic field,
the phase diagram of which has been recently obtained  
 by the DMRG approach \cite{okunishi03}.
In particular, a magnetization plateau \cite{oshikawa97,totsuka97,cabra97}
in the magnetization curve of the 
model was obtained at $1/3$ of the
full moment for strong enough anisotropy.  
This $m=1/3$ plateau state has a three-fold 
degenerate ground state corresponding to 
a spontaneous breaking of the translation symmetry.
It can be straightforwardly described in 
the regime of strong Ising anisotropy
$\Delta_{1,2} \gg J$ as the formation of an
up-up-down ($\uparrow \uparrow \downarrow$) structure
\cite{okunishi03}.
In this letter, we shall investigate the fate 
of this $m=1/3$ plateau phase upon application 
of a transverse staggered magnetic field
which tends to order 
the spins antiferromagnetically along the $x$-direction.
Within a bosonization approach,
we shall show that the competition between
the up-up-down order along the z-direction and the N{\'e}el order along
the $x$-direction induces a $T=0$ quantum critical
point which falls into the Z$_3$ universality class.
The critical behavior of the $x$ and $z$ components of the spin
at the transition will also be determined.

The low-energy theory corresponding to 
the lattice model (\ref{latticemodel}) can 
be derived using an Abelian bosonization approach
when $\Delta_{1,2}, h_s \ll J$.
To this end, we need the continuum description of the 
spin-1/2 operators of the Heisenberg chain 
in a magnetic field in terms of a bosonic field $\Phi$
and its dual field $\Theta$ 
\cite{oshikawa97,cabra97,totsuka97}:
\begin{eqnarray}
\frac{S_n^z}{a} &\sim&  S^z(x) = \frac{m}{2a} + 
\frac{\partial_x \Phi}{\sqrt{2\pi}}
+ \frac{A_1^z}{\pi a} \cos (\sqrt{2\pi} \Phi - 2k_F x)
\nonumber \\
\frac{S_n^x+iS_n^y}{a} &\sim& S^+(x) =
(-1)^{x/a}\frac{e^{i \sqrt{2\pi}\Theta}}{\pi a}\left[ A_0^{+} 
\right.  \nonumber \\
&+&\left.
A_1^{+} \cos(\sqrt{2\pi} \Phi - 2k_F x) +.. \right],
\label{eq:boson-operators}
\end{eqnarray}
where $x=na$ ($a$ being the lattice spacing) and $m/2$ 
is the average lattice magnetization $\langle S_n^z\rangle$. 
The  Fermi momentum $k_F$ is related to the magnetization
of the chain as
$k_F = \pi(1 - m)/2a$
\cite{oshikawa97,cabra97,totsuka97}.
In Eq. (\ref{eq:boson-operators}), 
the coefficients of the expansion
$A_{1}^z, A_{0,1}^{+}$ are non-universal 
constants which can be determined numerically 
for a given lattice model \cite{hikihara02}.
Within this low-energy 
approach,
the $m=1/3$ magnetization plateau phase of
the model  (\ref{latticemodel}) for $h_s =0$ is easily recovered.
In that case, the Fermi momentum is given by:
$k_F = \pi/3a$.
The bosonized representation (\ref{eq:boson-operators}),
yields
an umklapp process in the
Hamiltonian (\ref{latticemodel}) 
stemming from the $S^zS^z$ interacting terms,
leading to the following bosonized Hamiltonian:
\begin{eqnarray}
  \label{eq:ham-umklapp}
 {\cal H}_{\rm u} &=& \frac{v}{2} 
\int dx\left[ K (\partial_x \Theta)^2 + \frac{1}{K} (\partial_x
\Phi)^2\right] \nonumber \\
&-& \frac{\lambda}{\pi a}\int dx \; \cos 3 \sqrt{2\pi} \Phi ,
\end{eqnarray}
where the Luttinger parameters $v,K$
depend on the microscopic parameters of the lattice Hamiltonian
(\ref{latticemodel}) with $K$ decreasing as $\Delta_{1,2}$ increase. 
The low-energy effective Hamiltonian (\ref{eq:ham-umklapp}) thus
corresponds to the $\beta^2 = 18 \pi K$ sine-Gordon model. 
The sine-Gordon form of  
Eq. (\ref{eq:ham-umklapp}) can also be justified on 
symmetry grounds. Indeed,  translation by one lattice
site ${\cal T}_a$ is described by: 
$\Phi \rightarrow \Phi + \sqrt{2\pi}/3$ in 
the bosonization approach \cite{oshikawa97}, and 
$\cos 3 \sqrt{2\pi} \Phi$ is, among the operators left invariant by
this symmetry, the one with the lowest scaling dimension. 
The  field theory defined by Eq. (\ref{eq:ham-umklapp}) becomes massive 
when $K<4/9$. In that regime, which can be reached provided
$\Delta_1,\Delta_2>0$ are sufficiently large \cite{schulz-mit},
the bosonic field $\Phi$ is pinned on one of the minima of the
sine-Gordon model (\ref{eq:ham-umklapp}) i.e. 
$\langle \Phi \rangle = p \sqrt{2\pi}/3$ for $\lambda > 0$,
$p$ being integer, so that  a magnetization plateau at $m=1/3$ of the
saturation magnetization is formed.
Taking into account the compactified nature
of the bosonic field $\Phi$ ($\Phi \sim \Phi + \sqrt{2\pi}$)
in Eq. (\ref{eq:boson-operators}), we observe that
there are three non-equivalent ground states: 
$\langle \Phi \rangle = 0, \sqrt{2\pi}/3, 2 \sqrt{2\pi}/3$ that are
transformed into each other by a translation of one lattice site. 
This three-fold ground-state degeneracy 
is the thus the result of the spontaneous tripling of the unit cell in the
plateau phase, as can 
also be seen from the 
decomposition (\ref{eq:boson-operators}), which shows that
$\langle S_n^z\rangle$ oscillates with a period $3$
while $\langle S_n^x\rangle = 0$ in the plateau phase.
 
The low-lying excitations of this $m=1/3$ plateau phase are 
massive kinks and  antikinks interpolating between consecutive ground states, 
i.e. configurations of the field $\Phi(x)$ such that
$\Phi(+\infty)-\Phi(-\infty) = \sqrt{2\pi}/3$. These massive kink excitations
thus carry fractional $S^z$ quantum number:
$S^z = \int_{-\infty}^{\infty} dx
\; \partial_x \Phi/\sqrt{2\pi} = 1/3$.
All these properties are in agreement with 
 the Ising anisotropic limit: $\Delta_1,\Delta_2 \gg J, h_s$. 
In that limit, the ground state is three-fold degenerate
for a magnetization per site $\langle S_n^z \rangle = 1/6$ with an up
up down ($\uparrow \uparrow \downarrow$) structure invariant by a
translation of three lattice sites. Moreover, the low-lying
excitations of this structure are known to be massive domain walls
carrying  a fractional $S^z=1/3$ quantum number \cite{bak82,okunishi03}
in full agreement with the bosonization approach. Up to now, we have
been considering the case of a field tuned especially to the value
$H=H_c$ so that the state with $m=1/3$ is the ground state. If the
field starts to deviate from the special value $H=H_c$, it will
generate a chemical potential for the kinks and the antikinks $(H-H_c)
\partial_x \Phi$. However, if this chemical potential is smaller than
the gap to create kink excitations, no kink will appear in the ground
state, and the magnetization will remain at $m=1/3$. For larger
deviation of the magnetic field, a commensurate-incommensurate
transition will occur and the magnetization will deviate from
$m=1/3$. This is the origin of the presence of the magnetization
plateau for a finite interval of the magnetic field. 

Let us now investigate the stability of the $m=1/3$
plateau phase upon switching on the 
transverse staggered magnetic field $h_s$. 
Clearly, for $h_s\to \infty$ the spins will be
polarized along the $x$-direction and the $m=1/3$ 
magnetization plateau will be destroyed. 
The open question concerns the nature of the phase transition between these
two strong coupling massive ordered phases. We begin with the
case  $H=H_c$, and later on turn to $H\ne H_c$. 
Let us write down the bosonized Hamiltonian corresponding
to the model (\ref{latticemodel}) when $h_s \ll J$:
\begin{eqnarray}
  \label{eq:ham-potts}
  &{\cal H}& =   \frac{v}{2} \int dx\left[ (\partial_x \Theta)^2 
  + (\partial_x \Phi)^2\right]
  \\
  &+& \frac {h_s}{\pi a} \int dx \; \cos\sqrt{2\pi/K} \Theta 
  - \frac{\lambda}{\pi a}\int dx \; \cos\sqrt{18 \pi K} \Phi ,
  \nonumber
\end{eqnarray}
where we have rescaled the bosonic fields by the 
Luttinger parameter $K$.
The model (\ref{eq:ham-potts}) is
a generalization of the sine-Gordon model (\ref{eq:ham-umklapp}) 
obtained 
by adding a vertex operator which depends on the 
dual field $\Theta$.
The scaling dimensions of the two perturbations
in Eq. (\ref{eq:ham-potts}) are respectively 
$\Delta_{h_s} = 1/(2K)$ and
$\Delta_{\lambda} = 9K/2$.
For $1/4< K < 4/9$, a regime which can be  reached
owing to the presence of two 
independent couplings $\Delta_{1,2} > 0$ in Eq. (\ref{latticemodel}),
the two interacting terms are strongly relevant
perturbations so that a phase transition is expected.

Let us first give an argument for the existence of 
a phase transition for finite and
nonzero $h_s$.  
For $h_s=0$, 
i.e. the $m=1/3$ plateau phase,
the model (\ref{eq:ham-potts}) reduces 
to the $\beta^2 = 18 \pi K$ sine-Gordon model on the field $\Phi$
with massive solitons as elementary excitations.
The soliton-creating operators in this sine-Gordon model
with topological charge $m$ read as follows:
\begin{equation}
{\cal O}_m = \exp\left(i \frac{m}{3} \sqrt{2\pi/K} \; \Theta \right),
\label{solitons}
\end{equation}
which, in the path integral representation, 
creates a discontinuity of the field $\Phi$
($\Phi \rightarrow \Phi + m\sqrt{2\pi/K} /3$)
along a contour of integration (``Dirac string'')
which enters in the definition of the dual field $\Theta$ \cite{lukyanov01}.
Upon switching on a non-zero value of $h_s$, 
we notice that the perturbation $\cos(\sqrt{2\pi/K} \Theta)$
in Eq. (\ref{eq:ham-potts})
is locally mutual with the solitons (\ref{solitons}) of 
the  $\beta^2 = 18 \pi K$ 
sine-Gordon model (\ref{eq:ham-umklapp}) on the bosonic field $\Phi$
and can be interpreted as creating or annihilating
a packet of three elementary solitons.
The locality of the interaction term $\cos(\sqrt{2\pi/K} \Theta)$
with respect 
to the solitons (\ref{solitons}) of the sine-Gordon
model (\ref{eq:ham-umklapp}) implies that no phase transition
occurs  as an infinitesimal staggered magnetic field is switched on.
The $m=1/3$ plateau phase is thus stable and extends to a  
non-vanishing value of $h_s$.
Let us now make  similar considerations 
for the opposite regime ($h_s \gg \lambda$) 
of the model (\ref{eq:ham-potts}) where we are now
perturbing around 
the $\beta^2 = 2\pi/K$ sine-Gordon 
model on the dual field $\Theta$.
The ground state of the model (\ref{eq:ham-potts})
with $\lambda=0$ is characterized by $\langle S^z_n \rangle =0$
and $\langle (-1)^{n} S^x_n \rangle \ne 0$
as it can be easily seen from 
the decomposition (\ref{eq:boson-operators}).
In the same way as for the $m=1/3$ plateau phase,
this N{\'e}el ordered phase along the $x$-direction
is stable up to a strictly positive value of $\lambda$. 
The conclusions for the two limits considered here
are only compatible if a quantum phase transition occurs 
at a finite non-zero value of the staggered magnetic field $h_s^{*}$
between these two strong coupling massive phases.

The nature of the emerging quantum criticality  at $h_s^{*}$
can be elucidated by noting that 
the low-energy field theory (\ref{eq:ham-potts})
also describes the critical properties
of the two-dimensional classical
XY model with a three-fold symmetry-breaking field defined 
by the following lattice Hamiltonian \cite{jose77}:
\begin{equation}
{\cal H}_{Z_3} = - J \sum_{<{\bf r}, {\bf r}^{'}>}
\cos\left(\varphi_{{\bf r}} - \varphi_{{\bf r}^{'}}\right)
- h \sum_{{\bf r}} \cos\left(3 \; \varphi_{{\bf r}}\right),
\label{xyanisoham}
\end{equation}
$\varphi_{{\bf r}}$ being the angle of the unit-length
rotor at site ${\bf r}$ of a square lattice, and 
$h$ breaks the continuous O(2) symmetry
of the XY model down to a discrete one, Z$_3$. For $h\to +\infty$, the
model (\ref{xyanisoham}) reduces to a $Z_3$ clock model which is
equivalent to the three-state Potts model. 
More generally, since at zero temperature the term $h \cos 3 \varphi$ selects
$\varphi=-2\pi/3, 0, 2\pi/3$ as the possible ground states, 
the model (\ref{xyanisoham}) is expected to display a phase transition
as a function of temperature in the three-state Potts (Z$_3$) universality
class. 
This Z$_3$ criticality has indeed been confirmed numerically
by means of series analysis \cite{hamer80},
and exact diagonalizations on finite samples \cite{roomany80}. 
It has been known for some time \cite{wiegmann78} 
that the universal properties of the model (\ref{xyanisoham})
are captured by the Hamiltonian (\ref{eq:ham-potts}).
In particular, the first two terms in Eq. (\ref{eq:ham-potts})
constitute the continuum description of 
the O(2) XY model, with the cosine of the dual field 
in Eq. (\ref{eq:ham-potts}) creating and annihilating 
topological vortices of the field $\varphi_{{\bf r}}$. 
In addition, the vertex operator with coupling
constant $\lambda$ in Eq. (\ref{eq:ham-potts})
represents the continuum limit of the 
three-fold symmetry breaking perturbation.
The phase transition of the model
(\ref{latticemodel}) at $h_s^{*}$, separating the 
$m=1/3$ plateau phase from the N{\'e}el ordering along
the $x$-direction, should thus belongs to the Z$_3$ universality class.

Some insights of the physical properties
of the Z$_3$ quatum critical point
might be gained by considering
an analytic  approach \textit{\`a la} Luther-Emery
\cite{luther74}  of the generalized
sine-Gordon model~(\ref{eq:ham-potts}). 
At the special value $K=1/3$ and by fine-tuning
the coupling constants, 
the model (\ref{eq:ham-potts}) simplifies as follows
\begin{eqnarray}
{\cal H}_{\rm SDSG} &=& \frac{v}{2} \int dx \left[
\left(\partial_x \Phi\right)^2 +
\left(\partial_x \Theta\right)^2 \right]
\nonumber \\
&+& g\int dx \left[\cos\left(\sqrt{6\pi} \; \Phi\right)
+ \cos\left(\sqrt{6\pi} \; \Theta\right)\right],
\label{SDSG}
\end{eqnarray}               
which exhibits a self-dual $\Phi \leftrightarrow \Theta$
symmetry and will be referred in the following 
as the $\beta^2 = 6 \pi$ SDSG model.
It has been shown recently in a non-perturbative fashion that
the $\beta^2 = 6 \pi$  SDSG model (\ref{SDSG})
displays
an emerging Z$_3$ quantum criticality \cite{lgn02,delfino}. 
The universal properties of the model (\ref{SDSG}) can
be determined using 
the ultraviolet-infrared (UV-IR) transmutation
of some fields of the massless flow of the $\beta^2 = 6 \pi$ SDSG model 
which has been obtained from a Toulouse limit analysis \cite{lgn02}.
In particular, it has been shown in Ref. \onlinecite{lgn02} that 
the vertex operators 
$\exp(\pm i \sqrt{2\pi/3} \;\Phi)$ of the $\beta^2 = 6 \pi$ SDSG model
identify in the IR limit to the two Z$_3$ spin fields
$\sigma_1$ and $\sigma_2 = \sigma_1^{\dagger}$ with
scaling dimension $2/15$ and $q=\pm 1$ Z$_3$ charge
respectively \cite{cftbook}.
Using the continuum representation 
(\ref{eq:boson-operators}), we then predict the 
leading asymptotics 
at the Z$_3$ critical point
($h_s = h_s^{*}$)
of the spin-spin 
correlation function in the z-direction: 
\begin{eqnarray}
\langle \langle S^z \left(x\right) S^z \left(y\right) \rangle \rangle
&\equiv&  \langle \left[S^z \left(x\right) - m/2a \right]
\left[S^z \left(y\right) - m/2a \right] \rangle \nonumber\\
&\sim & \frac{\cos\left(2\pi\left(x-y\right)/3a\right)}{|x-y|^{4/15}}.
\label{prediction}
\end{eqnarray}
In the same way, it is possible to show that the 
operator $\cos(\sqrt{6\pi} \Theta)$ transmutes in the IR
limit to the energy operator $\epsilon$ of the three-state Potts model
with scaling dimension $4/5$ \cite{cftbook}.
We thus obtain from Eq. (\ref{eq:boson-operators}) the 
leading asymptotics of the 
spin-spin correlation function along the $x$-direction 
at the Z$_3$ critical point as:  
\begin{equation}
\langle \langle S^x \left(x\right) S^x \left(y\right) \rangle \rangle
\sim \frac{(-1)^{\left(x-y\right)/a}}{|x-y|^{8/5}}.
\label{predictionbis}
\end{equation}     
The spin-spin correlation functions 
(\ref{prediction},\ref{predictionbis}) thus display
an algebraic behavior with universal exponents characteristic 
of the Z$_3$ universality class. When the system is tuned away from
the self-dual point, we have to consider which relevant operators can
be generated. Both the operators $(\partial_x \Phi)^2$ and $\cos
\sqrt{6\pi}\Theta-\cos \sqrt{6\pi}\Phi$ are mapped onto the Z$_3$ thermal 
operator $\epsilon$ in the IR limit. Can another relevant operator be generated
besides the thermal one ? The entire content on the Z$_3$ Potts model
is known\cite{cftbook} and the thermal operator is the 
only relevant operator having zero
conformal spin and preserving the global $Z_3$ symmetry of the
model. 
The other relevant operators of the Z$_3$ Potts model
carry a non-zero conformal spin and only play a role
when one deviates from the center of the $m=1/3$ 
plateau phase as we will discuss below. 
As a result, all deviations from the self-dual point reduce here to
tuning the temperature in the Potts model away from the critical
temperature. The duality mapping $\Phi \leftrightarrow \Theta$,
$K\leftrightarrow 1/K$ can therefore be seen as the duality
transformation of the Potts model between the high and the low
temperature phase. The presence of the two independent coupling constants
$\Delta_2$ and $h_s$ of Eq. (\ref{latticemodel})
should be enough to reach the Z$_3$ critical point.

So far, we have only discussed the
case of a magnetic field $H=H_c$. 
If we now modify the magnetic field
to deviate slightly from the center of the $m=1/3$ plateau phase,
a Zeeman term ${\cal H}_Z = - (H - H_c) (K/2\pi)^{1/2}
\int dx \; \partial_x \Phi$
should be added to the low-energy Hamiltonian (\ref{eq:ham-potts}).
The resulting Hamiltonian describes a two-dimensional asymmetric
XY model in a three-fold breaking field:
\begin{equation}
{\cal H}_{\delta} = - J\sum_{{\bf r},\mu}
\cos(\varphi_{{\bf r}}-\varphi_{{\bf r} + {\bf e}_{\mu}}- \delta_{\mu})
- h \sum_{{\bf r}} \cos \left(3 \varphi_{{\bf r}}\right) ,
\label{asymxy}
\end{equation}
with $\delta_{\mu} \sim (H - H_c) {\bf e}_x \cdot {\bf e}_{\mu}, (\mu=x,y)$.
When $h \rightarrow \infty$, this model reduces to the
asymmetric three-state clock model \cite{ostlund} 
which displays commensurate, paramagnetic, and 
incommensurate phases \cite{ostlund,haldaneinc}.
Close to the Z$_3$ critical point, 
the low-energy effective Hamiltonian density 
corresponding to Eq. (\ref{asymxy})
is given by \cite{cardychir}:
\begin{equation}
{\cal H}_{\delta} = {\cal H}_{Z_3} + \delta  \Phi_{(2/5,7/5)}
+ \delta^{*}  \Phi_{(7/5,2/5)} 
+ \left(h_s - h_s^{*}\right)  \; \epsilon,
\label{chireff}
\end{equation}
where $\Phi_{(2/5,7/5)}$ and $ \Phi_{(7/5,2/5)}$
are Z$_3$ operators with 
scaling dimension $9/5$ and conformal spin $\pm 1$.
The model (\ref{chireff})
with $\delta^{*} = 0$  and $h_s = h_s^{*}$ 
is known to display critical behavior in the
chiral three-state Potts universality class
due to the presence of a non-Lorentz invariant perturbation \cite{cardychir}.
Such a chiral transition has been obtained recently 
in the phase diagram of a one-dimensional boson model with 
competing density-wave orders \cite{fendley}.
For $\delta = \delta^{*}$, there is a possibility 
of having a Lifshitz point between a chiral Z$_3$  
and a standard Z$_3$ critical lines.
In this respect, it will be interesting
to investigate numerically the phase diagram of 
the lattice model (\ref{latticemodel})
to further shed light on this issue.

\end{document}